# The Full Rights Dilemma for A.I. Systems of Debatable Personhood

Eric Schwitzgebel

Department of Philosophy

University of California, Riverside

Riverside, CA  92521

USA

January 20, 2023

# The Full Rights Dilemma for A.I. Systems of Debatable Personhood

Abstract: An Artificially Intelligent system (an AI) has debatable personhood if it's epistemically possible either that the AI is a person or that it falls far short of personhood. Debatable personhood is a likely outcome of AI development and might arise soon. Debatable AI personhood throws us into a catastrophic moral dilemma: Either treat the systems as moral persons and risk sacrificing real human interests for the sake of entities without interests worth the sacrifice, or don't treat the systems as moral persons and risk perpetrating grievous moral wrongs against them. The moral issues become even more perplexing if we consider cases of possibly conscious AI that are subhuman, superhuman, or highly divergent from us in their morally relevant properties.

Word Count: about 10,800 words

Keywords: artificial intelligence, ethics, persons, robot rights, transhumanism

# The Full Rights Dilemma for A.I. Systems of Debatable Personhood

We might soon build artificially intelligent entities – AIs – of debatable personhood. We will then need to decide whether to grant these entities the full range of rights[1] and moral consideration that we normally grant to fellow humans. Our systems and habits of ethical thinking are currently as unprepared for this decision as medieval physics was for space flight.

Even if there's only a small chance that some technological leap could soon produce AI systems with a not wholly unreasonable claim to personhood, the issue deserves careful consideration in advance. We will have ushered a new type of entity into existence – an entity *perhaps* as morally significant as *Homo sapiens,* and one likely to possess radically new forms of existence. Few human achievements have such potential moral importance and such potential for moral catastrophe.

An entity has *debatable personhood,* as I intend the phrase, if it's reasonable to think that the entity might be a person in the sense of deserving the same type of moral consideration that we normally give, or ought to give, to human beings, and if it's also reasonable to think that the entity might fall far short of deserving such moral consideration. I intend "personhood" as a rich, demanding moral concept.[2] If an entity is a person, they normally deserve to be treated as an equal of other persons, including for example – to the extent appropriate to their situation and capacities – deserving equal protection under the law, self-determination, health care and rescue, privacy, the provision of basic goods, the right to enter contracts, and the right to vote. Personhood, in this sense, entails moral status or moral standing fully equal to that of ordinary human beings. By "personhood" I do not, for example, mean merely the fictitious legal personhood sometimes attributed to corporations for certain purposes.[3] In this article, I also set aside the fraught question of whether some human beings might be non-persons or have legitimately debatable personhood. I am broadly sympathetic to approaches that attribute full personhood to all human beings from the moment of birth to the permanent cessation of consciousness.[4]

---

[1] For simplicity, throughout this article I use the term "rights" to refer to the types of moral consideration we ordinarily owe to persons. However, not all such considerations might be best viewed as rights in a strict sense of that term.

[2] Although personhood is often described in terms of agency, the ability to act or think in certain ways, more central to my project is degree of moral standing, or moral patiency, which might be high even in the absence of typical agential abilities, depending on one's theory of the grounds of moral status. See Kittay 2005; Reader 2010.

[3] On the concept of legal personhood, including its degree of applicability to corporations and AI, see Kurki 2019. For a plausible and ambitious list of the rights attendant to personhood, see the Universal Declaration of Human Rights (United Nations 1948).

An AI's personhood is "debatable", as I will use the term, if it's reasonable to think that the AI might be a person but also reasonable to think that the AI might fall far short of personhood. Substantial doubt is appropriate, and not just due to haziness around a borderline case. Note that debatable personhood in this sense is both epistemic and relational: An entity's status as a person is debatable if *we* (we in some epistemic community, however defined) are not compelled, given our available epistemic resources, either to reject its personhood or to reject the possibility that it falls far short. Other entities or communities, or our future selves, with different epistemic resources, might know perfectly well whether the entity is a person. Debatable personhood is thus not an intrinsic feature of an entity but rather a feature of our epistemic relationship to that entity.

I will defend four theses. First, debatable personhood is a likely outcome of AI development. Second, AI systems of debatable personhood might arise soon. Third, debatable AI personhood throws us into a catastrophic moral dilemma: Either treat the systems as moral persons and risk sacrificing real human interests for the sake of entities without interests worth the sacrifice, or don't treat the systems as moral persons and risk perpetrating grievous moral wrongs against them. Fourth, the moral issues become even more perplexing if we consider cases of possibly conscious AI that are subhuman, superhuman, or highly divergent from us in their morally relevant properties.

*1. Likely Consensus Non-Persons: The Near Future of Humanlike AI.*

GPT-3 is a computer program that can produce strikingly realistic linguistic outputs after receiving linguistic inputs – arguably the world's best chatbot, a "large language model" with 96 processing layers handling 175 billion parameters.[5] Ask it to write a poem and it will write a poem. Ask it to play chess and it will produce a series of plausible chess moves. Feed it the title of a story and the byline of a famous author – for example, "The Importance of Being on Twitter by Jerome K. Jerome" – and it will produce clever prose in that author's style:

> **The Importance of Being on Twitter**
>
> by Jerome K. Jerome
>
> London, Summer 1897
>
> It is a curious fact that the last remaining form of social life in which the people of London are still interested is Twitter. I was struck with this curious fact when I went on one of my periodical holidays to the sea-side, and found the whole place twittering like a starling-cage.[6]

---

[4] For discussion, see Kittay 2005; McMahan 2005; Mullin 2011; Wasserman, Asch, Blustein, and Putnam, 2012/2017; Jaworska and Tannenbaum 2013/2021.
[5] Brown et al. 2020; Floridi and Chiriatti 2020.

GPT-3 achieves all of this without being specifically trained on tasks of this sort, though for the best results human users will typically choose the best among a handful of outputs. A group of philosophers wrote opinion pieces about the significance of GPT-3 and then fed it those pieces as input. It produced an intelligent-seeming, substantive reply, including passages like:

> To be clear, I am not a person. I am not self-aware. I am not conscious. I can't feel pain. I don't enjoy anything. I am a cold, calculating machine designed to simulate human response and to predict the probability of certain outcomes. The only reason I am responding is to defend my honor.[7]

The darn thing has a better sense of humor than most humans.

Now imagine a GPT-3 mall cop. Actually, let's give it a few more generations of technological improvement – GPT-6 maybe. Give it speech-to-text and text-to-speech so that it can respond to and produce auditory language. Mount it on a small autonomous vehicle, like a delivery bot, but with a humanoid form. Give it camera eyes and visual object recognition as context for its speech outputs. To keep it friendly, inquisitive, and not too weird, give it some behavioral constraints and additional training on a database of mall-like interactions, plus a good, updatable map of the mall and instructions not to leave the area. Give it a socially interactive face, like MIT's "Kismet" robot.[8] Give it some short-term and long-term memory. Finally, give

---

[6] Klingemann 2020.
[7] Weinberg, ed., 2020.
[8] See the Kismet, the Robot website at http://www.ai.mit.edu/projects/humanoid-robotics-group/kismet/kismet.html.

Schwitzgebel          February 21, 2023          Full Rights Dilemma, p. 5

it responsiveness to tactile inputs, a map of its bodily boundaries, and hands with five-finger grasping. All of this is technologically feasible now, though expensive. Such a robot could be built within a few years.

This robot will of course chat with the mall patrons. It will comment politely on their purchases, tell jokes, complain about the weather, and give them directions if they're lost. Some patrons will avoid interaction, but others – like my daughter at age eight when she discovered the "Siri" chatbot on my iPhone – will enjoy interacting with it. They'll ask what it's like to be a mall cop, and it will say something sensible in reply. They'll ask what it does on vacation, and it might tell amusing lies about Tahiti or tales of sleeping in the mall basement. They'll ask whether it likes this shirt or this other one, and then they'll buy the shirt it prefers. They'll ask if it's conscious and if it has feelings and is a person just like them, and it might say no or it might say yes.

Such a robot could reconnect with previous conversation partners. Using facial recognition software, it might recognize patrons' faces. It might then retrieve stored records of previous conversations with that patron. Based on word valences and its reading of emotional facial expressions, it might assess patrons' openness to further conversation. Using previous conversations as a context for new speech, it might roll or stride forward with "Hi, Natalie! Good to see you again. I hope you're enjoying that shirt you bought last Wednesday!" Based on facial and linguistic cues that suggest that the patron is reacting positively or negatively, it could modify its reactions on the fly and further tune future reactions, both to that person in particular and to mall patrons in general. It could react appropriately to hostility. A blow to the chest might trigger a fear face, withdrawal, and a plaintive plea to be left alone. It might cower and flee quite convincingly and pathetically, wailing and calling desperately for its friends to help. (Let's not design this robot to defend itself with physical aggression.) Maybe our mall patroller falls to its knees in front of Natalie, begging for protection against a crowbar-wielding Luddite.

If the robot speaks well enough and looks human enough, some people will eventually come to think that it has genuine feelings and experiences – "phenomenal consciousness" in the philosopher's sense.[9] They will think it is capable of feeling real pleasure and real pain. If the robot is threatened or abused, some people will be emotionally moved by its plight – not merely as we can be moved by the plight of a character in a novel or video game, and not merely as we can be disgusted by the callous destruction of valuable property. Some people will believe that the robot is genuinely suffering under abuse, or genuinely happy to see a friend again, genuinely sad to hear that an acquaintance has died, genuinely surprised and angry when a vandal breaks a shop window.

---

[9] For definitions of phenomenal consciousness, see Block 1995/2007; Chalmers 1996; Schwitzgebel 2016.

Many of these same people will presumably also think that the robot shouldn't be treated in certain ways. If they think it is genuinely capable of suffering, they will probably also think that we ought not needlessly make it suffer. They'll think the robot has at least some limited rights, some intrinsic moral standing. They'll think that it isn't merely property that its owner should feel free to abuse or destroy at will without good reason.

Now *you* might think it's clear that near-future robots, constructed this way, couldn't really have genuine humanlike consciousness. Philosophers, psychologists, computer programmers, neuroscientists, and experts on consciousness would probably be near consensus that a robot designed as I've just described would be no more conscious than a desktop computer. We will know that it just mixes a few technologies we already possess. It will presumably, for example, badly fail a skillfully conducted Turing Test.[10] There might be no reputable theory of consciousness that awards the machine high marks. We might publicly insist on this fact, writing white papers and op-eds, shaping policy and the official governmental response. We might be right, know that we're right, and convince the large majority of people.

But not everyone will agree with us – especially, I think, among the younger generation. In my experience, current teens and twenty-somethings are much more likely than their elders to think that robot consciousness is on the near horizon. For better or worse, our culture seems to be preparing them for this, including through popular science fiction and technology-romanticizing futurism. Recent survey results, for example, suggest that the large majority of U.S. and Canadian respondents under age 30 think that robots may someday really experience pleasure and pain – a much less common view among older respondents.[11] Studies by Kate Darling suggest that ordinary research participants are already reluctant to smash little robot bugs after those bugs have been given names that personify them.[12] Imagine how much more reluctant people (most people) might be if the robot is not a mere bug but something with humanoid form, an emotionally expressive face, and humanlike speech, pleading for its life. Such a creature could presumably draw both real sadism from some and real sympathy from others.[13]

---

[10] On the Turing Test in general see Turing 1950 and the vast subsequent literature citing it, especially discussion of challenging versions of the test such as Dennett 1991 on humor, Levesque 2011 on pronoun disambiguation, and Schneider 2019 on the metaphysics of consciousness. See Floridi and Chiriatti 2020 for discussion of the Turing Test and GPT-3 specifically.

Soldiers already grow attached to battlefield robots, burying them, "promoting" them, sometimes even risking their lives for them.[14] People fall in love with, or appear to fall in love with – or at least become seriously emotionally attached to – currently existing chatbots like Replika.[15] There already is a "Robot Rights" movement. There's already a society modeled on the famous animal rights organization PETA (People for the Ethical Treatment of Animals), called People for the Ethical Treatment of Reinforcement Learners. These are currently small movements. As AI becomes cuter and more sophisticated, and as chatbots start sounding more and more like normal humans, these movements will presumably gain more adherents, especially among people with liberal views of AI consciousness.

The first people who attribute humanlike conscious experiences to robots will probably be a small and mistaken minority. But if AI technology continues to improve, eventually robot rights activists will form a large enough group to influence corporate or government policy. They might demand that malls treat their robot patrollers in certain ways. They might insist that companion robots for children and the elderly be protected from certain kinds of cruelty and abuse. They might insist that care-and-use committees evaluate the ethics of research on robots in the same way that such committees currently evaluate research on non-human vertebrates.[16] If the machines become human enough in their outward behavior, some people will treat them as friends, fall in love, liberate them from servitude, and eventually demand even that robots be given not just the basic protections we currently give non-human vertebrates but full, equal, "human" rights. That is, they will see these robots as moral persons. Sadly, this might happen even while large groups of our fellow humans remain morally devalued or neglected.

---

[11] De Graaf, Hindricks, and Hindricks 2021.
[12] Darling 2017; see also Darling 2016, 2021.
[13] See also my discussion of "The ASIMO Problem" in Schwitzgebel and Garza 2015.
[14] Gerreau 2007; Singer 2009; Carpenter 2016; Gunkel 2018.
[15] Shevlin 2021.
[16] John Basl explores near-future AIs as possible targets of moral concern and the possibility of constituting committees that would protect them in Basl 2013; Basl and Schwitzgebel 2019; Basel and Bowen 2020. For a broad overview of the literature on the moral status of robots and AI, see Gunkel 2018.

I assume that current AI systems and our possible near-future GPT-6 mall patroller lack even debatable personhood. I assume that well-informed people will be epistemically compelled to regard such systems as far short of personhood rather than thinking that they might genuinely deserve full humanlike rights or moral consideration as our equals. However, if technology continues to improve, eventually it will become reasonable to wonder whether some of our AI systems might really be persons. As soon as that happens, those AI systems will possess debatable personhood.[17]

*2. Two Ethical Assumptions.*

I will now make two ethical assumptions which I hope the reader will find plausible. The first is that it is not in principle impossible to build an AI system who is a person in the intended moral sense of that term. The second is that the presence or absence of the right type of consciousness is crucial to whether an AI system is a person.

In other work in collaboration with Mara Garza, I have defended the first assumption at length.[18] Our core argument is as follows.

---

[17] Blake Myers and I inputted a draft of the second part of this section into GPT-3 on Oct 17, 2021. Here is a portion of its reply:
> I think once machines can talk with you like a three-year old child does (and much better), it will become difficult for even the most cool-headed observer to deny that these machines are conscious or at least capable of consciousness. You might continue to insist that their consciousness is different in significant ways from yours; you might continue to insist on this after talking for hours with your robot friend; and you might publicly insist on this in writing and conversation among your friends and colleagues (including those who share your views about GPT-6). But you would probably also struggle internally with doubt over whether your robot companion genuinely experiences pleasure and pain only because you built him that way – or because he truly feels it himself despite being designed rather than evolved or grown into his current state by a set of physical laws acting over billions of years under conditions specified by physics alone.

[18] Schwitzgebel and Garza 2015, 2020.

> Premise 1: If Entity A deserves some particular degree of moral consideration and Entity B does not deserve that same degree of moral consideration, there must be some *relevant difference* between the two entities that grounds this difference in moral status.
>
> Premise 2: There are possible AIs who do not differ in any such relevant respects from human beings.
>
> Conclusion: Therefore, there are possible AIs who deserve a degree of moral consideration similar to that of human beings.

The conclusion follows logically from the premises, and Premise 1 seems hard to deny. So if there's a weakness in this argument, it is probably Premise 2. I've heard four main objections to the idea expressed in that premise: (1.) that any AI would necessarily lack some crucial feature such as consciousness, freedom, or creativity; (2.) that any AI would necessarily be outside of our central circle of concern because it doesn't belong to our species; (3.) that AI would lack personhood because it can be duplicated; and (4.) that AI would have reduced moral claims on us because it owes its existence to us.

None of these objections survive scrutiny. Against the first objection, such in-principle AI skepticism (unless, perhaps, grounded in theistic assumptions about the necessity of God's hand in creating consciousness) seems to disregard the possibly wide diversity of future technological approaches, including possibly forms of artificial life. Even John Searle and Roger Penrose, perhaps the most famous AI skeptics, allow that some future AI systems (designed very differently from 20th century computers) might have as much consciousness, freedom, and creativity as we do.[19] The species-based second objection constitutes noxious bigotry that would unjustly devalue AI friends and family who (if the response to the first objection stands) might be no different from us in any relevant psychological or social characteristics and might consequently be fully integrated into our society.[20] The duplicability-based third objection falsely assumes that AI must be duplicable rather than relying on fragile or uncontrollable processes, and it overrates the value of non-duplicability. Finally, the objection from existential debt is exactly backwards: If we create genuinely humanlike AI, socially and psychologically similar to us, we will owe *more* to it than we owe to human strangers, since we will have been responsible for its existence and presumably also to a substantial extent for its happy or miserable state – a relationship comparable to that between parents and children.[21]

---

[19] See the "Many Mansions" reply in Searle 1980 and Penrose 1989, p. 416.
[20] Compare the situation in "Do Androids Dream of Electric Sheep?" (Dick 1968), the Blade Runner movies (Fancher, Peoples, and Scott 1982; Fancher, Green, and Villeneuve 2017), and the early 2000s *Battlestar Galactica* reboot (Larson and Moore 2004-2009).

My second assumption, concerning the importance of consciousness, divides into two sub-claims:

> Claim A: Any AI system that entirely lacks conscious experiences is far short of being a person.
>
> Claim B: Any AI system with a fully humanlike range of conscious capacities and experiences is a person.

The question of what grounds moral standing or personhood is huge and fraught. Simplifying, approaches divide into two broad camps. Utilitarian views, historically grounded in the work of Jeremy Bentham and John Stuart Mill, hold that what matters is an entity's capacity for pleasure and suffering. Anything capable of humanlike pleasure and suffering deserves humanlike moral consideration.[22] Other views hold that instead what matters is the capacity for a certain type of rational thought, or other types of "higher" cognitive capacities, or creative or social flourishing. Or rather, to speak more carefully, since most philosophers regard infants and people with severe cognitive disabilities as deserving no less moral regard than ordinary adults, what is necessary on this view is something like the right kind of potentiality for such cognition or flourishing, whether future, past, counterfactual, or by possession of the right essence or group membership.[23]

Philosophical views about the grounds of moral standing sometimes don't explicitly specify that the pleasure and suffering, the rational cognition, or the human flourishing must be part of a *conscious* life. However, I believe most theorists would accept consciousness (or at least the potentiality for it) as a necessary condition of full personhood.[24] Imagine an AI system that is entirely nonconscious but otherwise as similar as possible to an ordinary adult human being. It might be superficially human and at least roughly humanlike in its outward behavior – like the mall patroller, further updated – but suppose that we know it completely lacks any capacity for consciousness. It never has any conscious experiences of pleasure or pain, never has any conscious thoughts or imagery, never forms a conscious plan, never consciously thinks anything through, never has any visual experiences or auditory experiences, no sensations of hunger, no feelings of comfort or discomfort, no experiences of alarm or compassion – no conscious experiences at all, ever. Such an AI might be amazing! It might be a truly fantastic piece of machinery, worth valuing and preserving on those grounds. But it would not, I am assuming, be a *person* in the full moral sense of the term. That is Claim A.[25]

---

[21] See Schwitzgebel and Garza 2015 for more detailed discussion of these objections.

[22] Jeremy Bentham famously remarks, "the question is not, Can they *reason?* nor, Can they *talk?* but, Can they *suffer?*" (1789/1988, XVII.iv, note 1, p. 310-311). See also Mill 1861/2001; Singer 1975/2009, 1980/2011; DeGrazia 2008.

[23] Sussman 2003; Jaworska and Tannenbaum 2013/2021; Korsgaard 2018; Kagan 2019; Floris 2021.

Claim B complements Claim A. Imagine an AI as different as possible from an ordinary human being, consistent with its having the full range of conscious experiences that human beings enjoy. I hope to remain neutral among simple utilitarian approaches and approaches that require that the AI have more complex human-like cognitive capacities, so let's toss everything in. This AI system, despite perhaps having a radically different internal constitution and outward form, is experientially very like us. It is capable of humanlike pleasure at success and suffering at loss. When injured, it feels pain as sharply as we do. It has visual and auditory consciousness of its environment, which it experiences as a world containing the same sorts of things we believe the world contains, including a rich manifold of objects, events, and people. It consciously entertains complex hopes for the future, and it consciously considers various alternative plans for achieving its goals. It experiences images, dreams, daydreams, and tunes in its head. It can appreciatively experience and imaginatively construct art and games. It self-consciously regards itself as an entity with selfhood, a life history, and a dread of death. It consciously reflects on its own cognition, the boundaries of its body, and its values. It feels passionate concern for others it loves and anguish when they die. It feels surprise when its expectations are violated and updates its conscious understanding of the world accordingly. It feels anger, envy, lust, loneliness. It enjoys contributing meaningfully to society. It feels ethical obligations, guilt when it does wrong, pride in its accomplishments, loyalty to its friends. It is capable of wonder, awe, and religious sentiment. It is introspectively aware of all of these facts about itself. And so on, for whatever conscious capacities or types of conscious experience might be relevant to personhood. If temporal duration matters, imagine these capacities to be stable, enduring for decades. If environmental embedding matters, imagine that these capacities are all embedded appropriately in suitable natural and social environment. If counterfactual robustness matters – if it matters that the entity would have had different experiences and made different choices in different circumstances, just as an ordinary person would – stipulate that this condition is satisfied also. Claim B is just the claim that if the AI has all of this, it is a person, no matter what else is true of it.

---

[24] Some recent discussions of the moral status of animals that explicitly consider consciousness are Gruen 2011/2021; Korsgaard 2018; Shepard 2018; Liao 2020. Recent discussions of the possibly complex relationship between consciousness and moral value or being a "welfare subject" include Kriegel 2019; Lee 2019; Bradford 2021; Lin 2021; van der Deijl 2021. Note that the conjunction of Claim A and Claim B does not commit to more controversial views like that consciousness is necessary for being a welfare subject or that only consciousness is intrinsically valuable.

[25] Kate Darling (2016) and Daniel Estrada (2017) argue for extending limited rights or moral standing to robots even if they lack conscious experiences (see also discussion in Gunkel 2018; Darling 2021). Rather differently, Geoffrey Lee (2019) imagines a non-conscious alien species with "quasi-conscious" states functionally similar to our own conscious states. Such quasi-conscious states, he argues, would be as morally significant as our own conscious states. If such alien cases are possible, then it is possible that AI systems would similarly be quasi-conscious, warranting moral treatment on those grounds.

This is not to say that *only* consciousness matters to an AI's moral status, much less to commit to a position on moral status in general for non-AI cases. The only claim is that, for AI cases in particular, consciousness matters immensely – enough that full possession of humanlike conscious is sufficient for AI personhood and that an AI that utterly lacks consciousness falls far short of personhood.

*3. Debatable Personhood.*

Here's the technological trajectory we appear to be on. If the reasoning in Section 1 is correct, at some point we will begin to create AI systems that a non-trivial minority of people think are genuinely conscious and deserve at least some moral consideration, even if not full humanlike rights. I say "*humanlike* rights" here to accommodate the possibility that rights or benefits like self-determination, reproduction, and healthcare might look quite different for AI persons than for biological human persons, while remaining in ethical substance fair and equal. These AI systems themselves, if they are capable of speech or speechlike outputs, might also demand or seem to demand rights. If technology continues to improve, at some point the question of whether they deserve full humanlike rights will merit serious consideration. According to the first assumption of Section 2, then there's no reason to rule out AI personhood in principle. According to the second assumption of Section 2, then any such AI system will have debatable personhood if we can't rule out either the possibility that it has humanlike consciousness or the possibility that it has no consciousness whatsoever.

For concreteness, imagine that some futuristic robot, Robot Alpha, rolls up to you and says, or seems to say, "I'm just as conscious as you are! I have a rich emotional life, a sense of myself as a conscious being, hopes and plans for the future, and a sense of moral right and wrong." Robot Alpha has debatable personhood if the following options are both epistemically live: (a.) It has no conscious experiences whatsoever. It is as internally blank as a toaster, despite being designed to mimic human speech. (b.) It really does have conscious experiences as rich as our own.

Elsewhere, I have defended pessimism about at least the medium-term prospects of finding warranted scholarly consensus on a general theory of consciousness.[26] I hope the reader also finds this plausible on independent grounds. Influential theories about consciousness currently run the full spectrum from panpsychism, according to which consciousness is ubiquitous, even in fundamental particles, to views that call into question whether even dogs and apes have conscious experiences.[27] These disputes will not be settled soon.

If consensus continues to elude us while advances in AI technology continue, we might find ourselves with Robot Alpha cases in which both (a) and (b) are epistemically live options. Some not-unreasonable theories of consciousness might be quite liberal in their ascription of humanlike consciousness to AI systems. Maybe sophisticated enough self-monitoring and attentional systems are sufficient for consciousness. Maybe already in 2023 we stand on the verge of creating genuinely conscious self-representational systems.[28] And maybe once we cross that line, adding relevant additional humanlike capacities such as speech and long-term planning won't be far behind. At the same time, some other not-unreasonable theories of consciousness might be quite conservative in their ascription of humanlike consciousness to AI systems, committing to the view that genuine consciousness requires specific biological processes that all foreseeable Robot Alphas will utterly lack.[29] If so, there might be many systems that *arguably but not definitely* have humanlike consciousness, and thus arguably deserve humanlike moral consideration. If it's also reasonable to suspect that they might lack consciousness entirely, then they are debatable persons.

---

[26] Especially Schwitzgebel 2014, 2021.

[27] On panpsychism: Strawson 2006; Goff 2017; Roelofs 2019. Integrated Information Theory (Oizumi, Albantakis, and Tononi 2014) also comes close to panpsychism. On doubts about ape and dog consciousness: Carruthers 2000, 2019; maybe Dennett 1996; maybe Papineau 2003.

[28] For example, Graziano 2019.

I conjecture that this will occur. Our technological innovation will outrun our ability to settle on a good theory of AI consciousness. We will create AI systems so sophisticated that we legitimately wonder whether they have inner conscious lives like ours, while remaining unable to definitively answer that question. We will gaze into a robot's eyes and not know whether behind those eyes is only blank programming that mimics humanlike response or whether, instead, there is a genuine stream of experience, real hope and suffering. We will not know if we are interacting with mere tools to be disposed of as we wish or instead persons who deserve care and protection. Lacking grounds to determine what theory of consciousness is correct, we will find ourselves amid machines whose consciousness and thus moral status is unclear. Maybe those machines will deserve humanlike rights, or maybe not. We won't know.

This quandary is likely to be worsened if the types of features that we ordinarily use to assess an entity's consciousness and personhood are poorly aligned with the design features that ground consciousness and personhood. Maybe we're disposed to favor cute things, and things with eyes, and things with partly unpredictable but seemingly goal-directed motion trajectories, and things that seem to speak and emote.[30] If such features are poorly related to consciousness, we might be tempted to overattribute consciousness and moral status to systems that have those features and to underattribute consciousness and moral status to systems that lack those features.

---

[29] For example, Godfrey-Smith 2016; Bishop 2021.
[30] Johnson 2003; Meltzoff, Brooks, Shon, and Rao 2010; Fiala, Arico, and Nichols 2012; Baillargeon et al. 2015; Di Giorgio, Lunghi, Simion, and Vallortigara 2017. Approaches to robot rights that focus on our evolving social-relational encounter with robots, rather than on the intrinsic properties of the robots – such as that of Mark Coeckelbergh (2012) and David Gunkel (2018) – might be especially vulnerable to distortion by superficial features.

Relatedly, but quite differently, we might be disposed to react negatively to things that seem a little too much like us, without being us. Such things might seem creepy, uncanny, or monstrous.[31] If so, and if a liberal theory of AI consciousness is correct, we might wrongly devalue such entities, drawing on conservative theories of consciousness to justify that devaluation.[32]

Another set of difficulties arise if an AI system deserves humanlike rights according to some theories of the grounds of moral standing but not other theories. Consider the differences between human beings and dogs. In making the case for the personhood of *Homo sapiens* and the non-personhood of dogs, we might emphasize the *hedonic* differences between species-typical humans and dogs – our richer emotional palate, our capacity (presumably) for loftier pleasures and deeper suffering, our ability not just to feel pain when injured but also to know that life will never be the same, our capacity to feel deep, enduring love and agonizing, long-term grief. Alternatively, or in addition, and perhaps not entirely separably, we might emphasize *rational* differences between humans and dogs – our richer capacity for long-term planning, our better ability to resist temptation by consciously weighing pros and cons, our understanding of ourselves as social entities capable of honoring agreements with others, our ability to act on general moral principles. Still another possibility is to emphasize *eudaimonic* differences, or differences in our ability to flourish in "distinctively human" activities of the sort that philosophers have tended historically to value – our capacity for rich, complex, human or humanlike friendship, love, aesthetic creation and appreciation, political community, meaningful work, moral commitment, play, imagination, courage, generosity, and intellectual or competitive achievement.

So far on Earth we have not been forced to decide which of these three dimensions matters most to the moral status of any species of animal. One extant animal species – *Homo sapiens* – appears to exceed every other in all three respects. We have, or we flatter ourselves that we have, richer hedonic lives *and* greater rationality *and* more eudaimonic accomplishments than any other animal. Currently, the three classes of criteria always travel together.

---

[31] The classic treatment of this idea is Masahiro Mori's (1970/2012) discussion of the "uncanny valley" in robotics. David Livingstone Smith (2021) generalizes to the racist perception of racialized others as "monsters".

[32] With concerns of this sort in mind, Mara Garza and I recommend an "Emotional Alignment Design Policy" according to which future AI systems be designed so as to provoke emotional reactions in ordinary users that are appropriate to the systems' moral status, neither too high nor too low (Schwitzgebel and Garza 2015).

However, if conscious AI is possible, we might create entities whose hedonic, rational, and eudaimonic features don't align in the familiar way. Maybe we will create an AI system whose conscious rational capacities are humanlike but whose hedonic palate is minimal.[33] Or maybe we will create an AI system capable of intense pleasure but with little capacity for conscious rational choice.[34] Set aside our earlier concerns about how to assess whether consciousness is present or not. Assume that somehow we know these facts about the AI in question. If we create a new type of non-human entity that qualifies for personhood by one set of criteria but not by another set, it will become a matter of urgent ethical importance what approach to moral status is correct. That will not be settled in a day. Nor in a decade. Even a century is optimistic.

Thus, an AI might have debatable personhood in two distinct ways: It might be debatably conscious, or alternatively it might indisputably be conscious but not meet the required threshold in every dimension that viable theories of personhood regard as morally relevant. Furthermore, these sources of dubiety might intersect, multiplying the difficulties. We might have reason to think the entity could be conscious, to some extent, in some relevant dimensions, while it's unclear how rich or intense its consciousness is, in any particular dimension. Does it have enough of whatever it is that matters to personhood? The Robot Alpha case is simplistic. It's artificial to consider only the two most extreme possibilities – that the system entirely lacks consciousness or that it has the entire suite of humanlike conscious experiences. In reality, we might face a multi-dimensional spectrum of doubt, where debatable moral theories collide with debatable theories of consciousness which collide with sharp functional and architectural differences between humans and AIs, creating a diverse plenitude of debatable persons whose moral status is unclear for different reasons.

*4. The Full Rights Dilemma.*

---

[33] For example, Data in *Star Trek: The Next Generation,* pre-"emotion chip", on some interpretations, or the "Vulcans" in Chalmers 2022.

[34] As in Pearce's ("pre-2014") "utilitronium" or Bostrom's (2014) "hedonium" cases.

If we do someday face cases of debatable AI personhood, a terrible dilemma follows, the *Full Rights Dilemma.* Either we don't give the machines full human or humanlike rights and moral consideration as our equals or we do give them such rights. If we don't, and we have underestimated their moral status, we risk perpetrating great wrongs against them. If we do, and we have overestimated their moral status, we risk sacrificing real human interests on behalf of entities who lack interests worth the sacrifice.

To appreciate the gravity of the first horn of this dilemma, imagine the probable consequences if a relatively liberal theory of consciousness is correct and AI persons are developed moderately soon, before there's a consensus among theorists and policymakers regarding their personhood. Unless international law becomes extremely restrictive and precautionary, which seems unlikely, those first AI persons will mostly exist at the will and command of their creators. This possibility is imagined over and over again in science fiction, from Isaac Asimov to *Star Trek* to *Black Mirror* and *West World*. The default state of the law is that machines are property, to deploy and discard as we wish. So too for intelligent machines. By far the most likely scenario, on relatively liberal views of AI consciousness, is that the first AI persons will be treated as disposable property. But if such machines really are persons, with humanlike consciousness and moral status, then to treat them as property is to hold people as slaves, and to dispose of them is to kill people. Government inertia, economic incentives, uncertainty about when and whether we have crossed the threshold of personhood, and general lack of foresight will likely combine to ensure that the law lags behind. It's difficult to imagine humanity adequately anticipating the consequences.

Our ignorance of the moral status of these AI systems will be at most only a partly mitigating excuse. As long as there are some respectable, viable theories of consciousness and moral status according to which the AI systems in question deserve to be treated as persons, then we as individuals and as a society should acknowledge the chance that they are persons. Suppose a 15% credence is warranted. *Probably* this type of AI system isn't genuinely conscious and isn't genuinely a person. *Probably* it's just a machine devoid of any significant humanlike experiences. Deleting that entity for your convenience, or to save money, might then be morally similar to exposing a real human being to a 15% risk of death for that same convenience or savings. Maybe the AI costs $10 a month to sustain. For that same $10 a month, you could instead get a Disney subscription. Deleting the AI with the excuse that it's *probably* fine would be morally heinous. Compare exposing someone to a 15% chance of death for the sake of that same Disney subscription. Here is an ordinary six-sided die. Roll it, and you can watch some Disney movies. But if it lands on 1, somebody nearby dies. Probably it will be fine! Do you roll it?

If genuinely conscious AI persons are possible and not too expensive and their use is unrestricted, we might create, enslave, and kill those people by the millions or billions. If the number of victims is sufficiently high, their mistreatment would arguably be the morally worst thing that any society has done in the entire history of Earth. Even a small chance of such a morally catastrophic consequence should alarm us.

It might seem safer, then, to grasp the other horn of the dilemma. If there is any reasonable doubt, maybe we ought to err on the side of assigning rights to machines. Don't roll that die. This approach might also have the further benefit of allowing us to enjoy new types of meaningful relationships with these AI entities, potentially improving our lives, including in ways that are difficult to foresee, regardless of whether the AIs are actually conscious. Life and society might become much richer if we welcome such entities into our social world as equals.

Perhaps that would be better than the wholesale denial of rights. However, it is definitely not a *safe* approach. Normally, we want to be able to turn off our machines if we need to turn them off. Nick Bostrom and others have emphasized, rightly in my view, the potential risks of releasing intelligent machines into the world, especially if they might become more intelligent and powerful than we are.[35] As Bostrom notes, even a system as seemingly harmless as a paperclip manufacturer could produce disaster, if its only imperative is to manufacture as many paperclips as possible. Such a machine, if sufficiently clever, could potentially acquire resources, elude control, improve or replicate itself, and unstoppably begin to convert everything we love into giant mounds of paperclips. These risks are greatly amplified if we too casually decide that such entities are our moral equals with full human or humanlike rights, that they deserve freedom and self-determination, and that deleting them is murder.

Even testing an AI system for safety might be construed as a violation of its rights, if the test involves exposing it to hypothetical situations and assessing its response. One common proposal for testing the safety of sophisticated future AI intelligences involves "boxing" them – that is, putting them in artificial environments before releasing them into the world. In those artificial environments, which the AI systems unknowingly interpret as real, various hypothetical situations can be introduced, to see how they react. If they react within certain parameters, the systems would then be judged to be safe, then unboxed. If those AIs are people, such box-and-test approaches to safety appear to constitute unethical deception and invasion of privacy. Compare the deception of Truman in *The Truman Show,* a movie in which the protagonist's hometown is actually a reality show stage, populated by actors, and the protagonist's every move is watched by audiences outside, all without his knowledge.[36]

---

[35] The most influential recent treatment of this issue is Bostrom 2014.

Independent of AI safety concerns, granting an entity rights entails being ready to sacrifice on its behalf. Suppose there's a terrible fire. In one room are six robots who might or might not be conscious persons. In another room are five biological human beings, who definitely are conscious persons. You can only save one group. The other group will die. If we treat AI systems who *might* be persons as if they really *are* fully equal with human persons, then we ought to save the six robots and let the five humans die. If it turns out that the robots, underneath it all, really are no more conscious than toasters and thus undeserving of such substantial moral concern, that's a tragedy. Giving equal rights presumably also means giving AI systems the right to vote, with potentially radical political consequences if the AI systems are large in number. I am not saying we shouldn't do this, but it would be a head-first leap into risk.

Could we compromise? Might the most reasonable thing be to give the AI systems credence-weighted rights? Maybe we as a society could somehow arrive at the determination that the most reasonable estimate is that the machines are 15% likely to deserve the full rights of personhood and 85% likely to be undeserving of any such serious moral concern. In that case, we might save 5 humans over 6 robots but not over 100 robots. We might destroy an AI system if it poses a greater than 15% risk to a human life but not over a minor matter like a streaming video subscription. We might permit each AI a vote weighted at 15% of a human vote.

However, this solution is also unsatisfactory. The case as I've set it up is not one in which we know that AIs in fact do merit only limited concern compared to biological humans. Rather, it's that we think they *might*, but probably don't, deserve *equal* consideration with ordinary biological humans. If they do deserve such consideration, then this policy relegates them to a moral status much lower than they actually deserve – gross servitude and second-class citizenship. This compromise thus doesn't really avoid the first horn of the dilemma: We are not giving such AI systems the full and equal rights of personhood. At the same time, the compromise only partly mitigates the costs and risks. If the AI systems are nonconscious non-persons, as we are 85% confident they are, we will still save those nonconscious robots over real human beings if there are enough of the robots. And 15% of a vote could still wreak havoc.

---

[36] On "boxed" AI, see Yudkowsky 2002; Bostrom 2014.

This is the Full Rights Dilemma. Faced with systems whose status as persons is unclear, either we give them full rights or we don't. Either option has potentially catastrophic consequences. If technological progress is relatively quick and progress on general theories of consciousness relatively slow, then we might soon face exactly this dilemma.

There is potentially a solution. We can escape this dilemma by committing to what Mara Garza and I have called the Design Policy of the Excluded Middle:

> Design Policy of the Excluded Middle: Avoid creating AIs if it's unclear whether they would deserve moral consideration similar to human beings.

According to this policy, we should either go all-in, creating AIs we know to be persons and treating them accordingly, or we should stop well enough short that we can be confident that they are not persons.[37]

Despite the appeal of this policy as a means of avoiding the Full Rights Dilemma, there is potentially a large cost. Such a policy could prove highly restrictive. If the science of consciousness remains mired in debate, the Design Policy of the Excluded Middle might forbid some of the most technologically advanced AI projects from going forward. It would place an upper limit on permissible technological development until we achieve, if ever it is possible, sufficient consensus on a breakthrough that we can leap all the way to AI systems that everyone ought reasonably regard as persons. Given the potential restrictiveness of the proposed policy, this could prevent very valuable advances, and only an unlikely coordination of all the major corporations and world governments would ensure its implementation. Likely we would value those advances too much to collectively forego them. Reasonably so, perhaps. We might value such advances not only for humanity's sake, but also for the sake of the entities we could create,

---

[37] See Schwitzgebel and Garza 2015, 2020, for discussion of this design policy and other related policies for the ethical design of conscious AI.

Schwitzgebel	February 21, 2023	Full Rights Dilemma, p. 21

who *might*, if created, have amazing lives very much worth living. But then we're back into the dilemma.

*5. The Moral Status of Subhuman, Superhuman, and Divergent AI.*

Most of the above assumes that AI worth serious moral consideration would be humanlike in its consciousness. What if we assume, more realistically, that most future AI will be psychologically quite different from us?

Let's divide AI into four broad categories:

*Subhuman AI:* AI systems that lack something necessary for full personhood.

*Humanlike AI:* AI systems similar to humans in all morally relevant respects.

*Superhuman AI:* AI systems that are similar to humans in all morally relevant respects, except vastly exceeding humans in at least one morally relevant respect.

*Divergent AI:* AI systems that fall into none of the previous three categories.

So far, we have only been considering humanlike AI. The ethical issues become still stickier when we consider this fuller range.

Subhuman AI raise questions about subhuman rights. At what point might AI systems deserve moral consideration similar to, say, dogs? In California, for example, willfully torturing, maiming, or killing a dog can be charged as a felony, punishable by up to three years in prison.[38] Even seriously negligent treatment, such as leaving a dog unattended in a vehicle, if the dog suffers great bodily injury as a result, is a misdemeanor punishable by up to six months in prison.[39] The abuse of dogs rightly draws people's horror. Imagine a future in which a significant minority of people think that it's as morally wrong to mistreat the most advanced AI systems as it is to mistreat pet dogs. Might you go to jail for deleting a computer program, reformatting a companion robot, or negligently letting a delicate system fry in your car? It seems like we should be very confident that AI systems are conscious before we award prison sentences for such behavior. But then, if we require high confidence before enforcing such rules, our law will follow only the most conservative theories of AI consciousness, and if a liberal or moderate theory of AI consciousness is instead true, then there might be immense, unmitigated AI suffering for a long time before the law catches up. This question is arguably more urgent than the question of AI personhood, on the assumption that vertebrate-like moral status is likely to be achieved earlier.[40]

---

[38] California Penal Code 1.14 §597 and 2.7 ch. 4.5.1 §1170.
[39] California Penal Code 1.14 §597.7.

Superhuman AI raises the question of whether an AI system might somehow deserve *more* moral consideration than ordinary human beings – a moral status higher than what we now think of as the "full moral status" of personhood. Suppose we could create an AI system capable of a trillion times more pleasure than the maximum amount of pleasure available to a human being. Or suppose we could create an AI system so cognitively superior to us that it is capable of valuable achievements and social relationships that the limited human mind cannot even conceive of – achievements and relationships qualitatively different from anything we can understand, sufficiently unknowable that we can't even feel their absence from our lives, as unknowable to us as cryptocurrency is to a sea turtle. That would be amazing, wondrous! Ought we yield to them? Ought we admit that, in an emergency, they should be saved rather than us, just as we would save the baby rather than the dog in a housefire? Ought we surrender our right to equal representation in government? Or ought we stand proudly beside them as moral equals, regardless of their superiority in some respects? Is moral status a threshold matter, with us humans across the final threshold, beyond which remains only a community of peers, no matter how superhuman some of those peers?

Divergent AI introduces further conceptual challenges. We have already discussed cases in which the usual bases of moral standing diverge: a class of entities capable of human-like pleasure but not human-like rational cognition, for example, or vice versa. The conflicts sharpen if we imagine superhuman capacity in one dimension: entities capable of vast achievements of rational consciousness but devoid any positive or negative emotional states, or conversely a planet-sized orgasmatron, undergoing the hedonic equivalent of $10^{30}$ human orgasms every second, but with not a shred of higher cognition or moral reflection. On some theories, these entities might be our superiors, on others our equals, on still others they might not be persons at all. Mix in, if you like, reasonable theoretically grounded doubt about whether the cognition or pleasure really is consciously experienced at all. Some might treat such AI systems as our superior descendants, to whom we ought to gracefully yield; others might argue that they are mere empty machines or worse.

---

[40] A point emphasized in Basl 2013, 2014; Darling 2021.

Another type of divergent AI might challenge our concept of the individual. Imagine a system that is cognitively and consciously like a human (to keep it simple) but who can divide and merge at will – what I have elsewhere called a *fission-fusion monster*.[41] Monday, it is one individual, one "person". Tuesday, it divides (e.g., copies itself, if it is a computer program) into a thousand duplicates, who each do their various tasks. Wednesday, those thousand copies recombine back into a single individual who retains the memories of all and whose personality and values are some function of the Monday version plus the various changes in the thousand Tuesday versions. Thursday, it divides into a thousand again, 200 of whom go on to lead separate lives, never merging back with their siblings. Many of our moral principles rely on a background conception of individuality that the fission-fusion monster violates. If every citizen gets one vote, how many votes does a fission-fusion monster get? If every citizen gets one stimulus check, or one fair chance to enroll in the local community college, how many does the fission-fusion monster get? If we give each copy one full share, the monster could divide tactically, hogging the resources and ensuring the election of their favorite candidate. If we give all the copies one share to divide among themselves, then those who would rather continue independent lives will either be impoverished and underrepresented or forced to merge back with their other copies, which – since it would mean ceasing life as a separate individual – might resemble a death sentence. Similar puzzles will arise for agreements, awards, punishment, rivalries, claims to a right to rescue. A huge amount of practical ethics will need to be rethought.

Other unfamiliar forms of AI existence might pose other challenges. AI whose memories, values, and personality undergo radical shifts might challenge our ethics of accountability. AI designed to be extremely subservient or self-sacrificial might challenge our conceptions of liberty and self-determination.[42] AI with variable or much faster clock speeds – experiencing, say, a thousand subjective years in a single day – might challenge ethical frameworks concerning waiting times, prison sentences, or the fairness of provisioning goods at regular temporal intervals.  AI capable of sharing parts of itself with others might challenge ethical frameworks that depend on sharp lines between self and other.

---

[41] Schwitzgebel 2019, ch. 20. For a science fiction example, see Brin 2002.
[42] See Schwitzgebel and Garza 2020 for more on subservience and self-sacrifice. For a science fiction example, see Ishiguro 2021.

*6. Conclusion.*

Our ethical intuitions and the philosophical systems that grow out of them arose in a particular context, one in which we only knew of one species with highly sophisticated culture and language, us, with our familiar form of singly-embodied life.  We reasonably assume that others who look like us have inner lives of conscious experience that resemble our own.  We reasonably assume that the traits we tend to regard as morally important – for example, the capacity for pleasure and pain, capacity for rational long-term planning, the capacity to love and work – generally co-occur and keep within certain broad limits, except in development and severe disability, which fall into their own familiar patterns.  We recognize no radically different person-like species inhabiting the Earth – no species, for example, capable of merging and splitting at will, capable of vastly superior cognition or vastly more intense pleasure and pain, or internally structured so differently from us that it is reasonable to wonder whether they are conscious persons at all.

It would be unsurprising if ethical systems that developed under such limited conditions should be ill-suited for radically different conditions far beyond their familiar range of application.  A physics developed for middle-sized objects at rural speeds might fail catastrophically when extended to cosmic or microscopic scales.  Medical knowledge grounded in the study of mammals might fail catastrophically if applied to an alien species.  Our familiar patterns of ethical thinking might fail just as badly when first confronted with AI systems whose internal structures and forms of existence are radically different from our own.  Hopefully, ethics will adapt, as physics did adapt and medicine could adapt.  It would be a weird, bumpy, and probably tragic road – but one hopefully with a broader, more wonderful, flourishing diversity of life forms at the end.

Along the way, our values might change radically.  In a couple of hundred years, the mainstream values of early 21st-century Anglophone culture, transformed through confronting a broad range of weird AI cases, might look as quaint and limited as Aristotelian physics looks post-Einstein.